\documentclass{nature_SI}
\usepackage{graphicx}
\usepackage{amssymb}
\usepackage{mathtools}

\title{Graphene mechanical pixels for Interferometric MOdulator Displays (GIMOD)}

\author{Santiago J. Cartamil-Bueno$^{1,*}$, Dejan Davidovikj$^{2}$, Alba Centeno$^{3}$, Amaia Zurutuza$^{3}$, Herre S.J. van der Zant$^{2}$, Peter G. Steeneken$^{2}$ \& Samer Houri$^{2,*}$\footnote[3]{Current affiliation: NTT Basic Research Laboratories, NTT Corporation, 3-1 Morinosato-Wakamiya, Atsugi, Kanagawa 243-0198, Japan}}
\begin{document}
\maketitle

\begin{affiliations}
	\item Advanced Microelectronic Center Aachen (AMICA), AMO GmbH, Otto-Blumenthal-Str. 25, 52074 Aachen, Germany
	\item Kavli Institute of Nanoscience, Delft University of Technology, Lorentzweg 1, 2628CJ, Delft, The Netherlands
	\item Graphenea SA, 20018 Donostia-San Sebasti\'an, Spain\\
$^*$Corresponding authors: cartamil@amo.de., Houri.Samer@lab.ntt.co.jp.
\end{affiliations}
\nointerlineskip

\noindent
\begin{abstract}
Graphene, the carbon monolayer and 2D allotrope of graphite, has the potential to impact technology with a wide range of applications such as optical modulators for high-speed communications~\cite{Bonaccorso2010,Ferrari2014,Yu2015,Sun2016}. In contrast to modulation devices that rely on plasmonic or electronic effects, MEMS-based modulators can have wider tuning ranges albeit at a lower operating frequency. These properties make electro-optic mechanical modulators ideal for reflective-type display technologies as has been demonstrated previously with SiN membranes in Interferometric MOdulator Displays (IMODs)~\cite{Chan2017}.
Despite their low-power consumption and performance in bright environments, IMODs suffer from low frame rates and limited color gamut.
Double-layer graphene (DLG) membranes grown by chemical vapor deposition (CVD) can also recreate the interference effect like in IMODs as proven with drumheads displaying Newton's rings~\cite{Cartamil-Bueno2016}.
Here, we report on the electro-optical response of CVD DLG mechanical pixels by measuring the change in wavelength-dependent reflectance of a suspended graphene drumhead as a function of electrical gating. We use a spectrometer to measure the wavelength spectrum at different voltages, and find a good agreement with a model based on light interference.
Moreover, to verify that gas compression effects do not play an important role, we use a stroboscopic illumination technique to study the electro-optic response of these graphene pixels at frequencies up to 400~Hz. Based on these findings, we demonstrate a continuous full-spectrum reflective-type pixel technology with a Graphene Interferometric MOdulator Display (GIMOD) prototype of 2500~pixels per inch (ppi) equivalent to more than 12K resolution.
\end{abstract}

\begin{figure}[h]
	\centering
	\graphicspath{{Figures/}}
	\includegraphics[width=177.8mm]{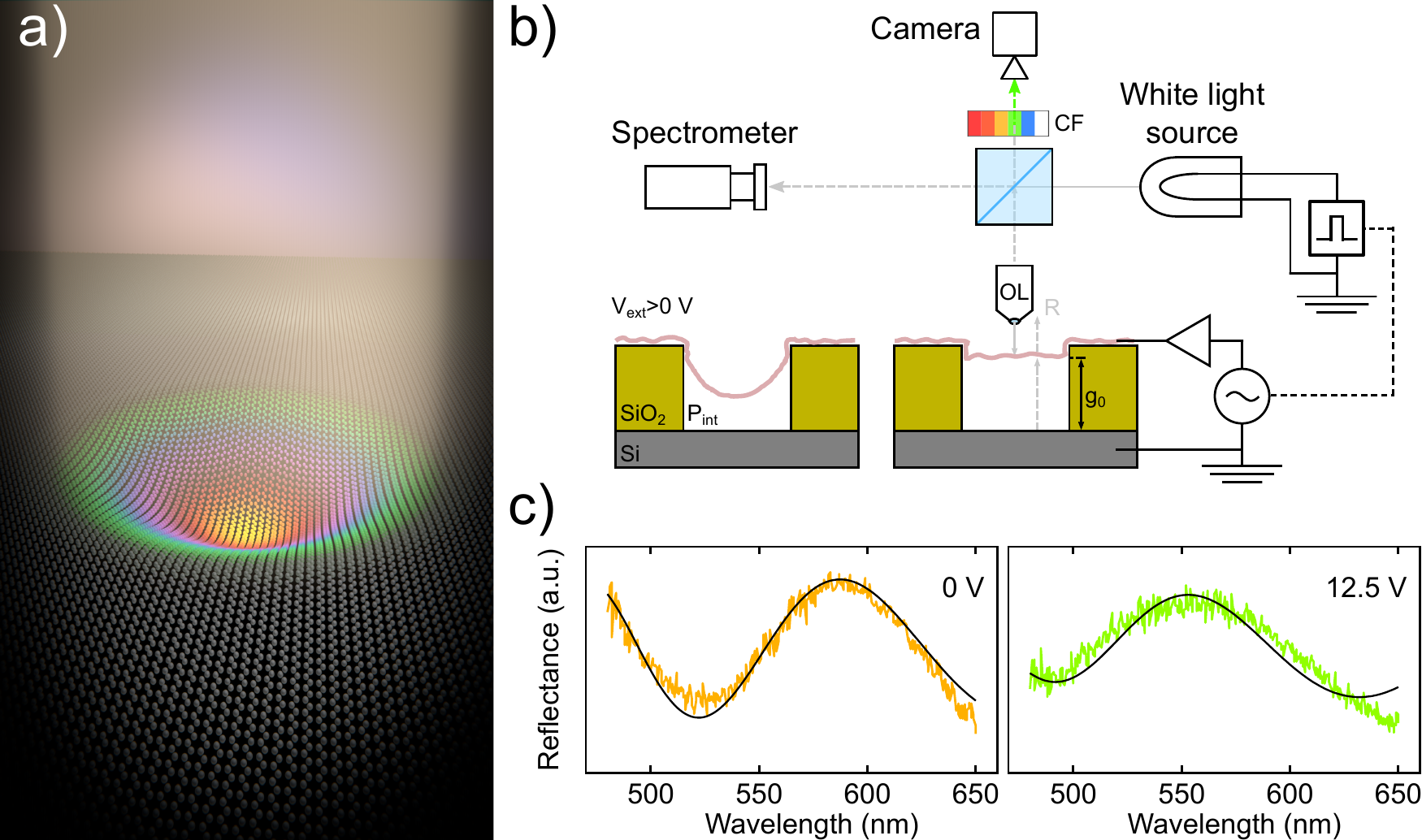}
	\caption{Electro-optical measurements with a colorimetry setup of a graphene drumhead of 20~$\mu$m in diameter. a) Artist impression of a deformed CVD DLG drumhead showing Newton's rings upon white light illumination. b) The membrane can be actuated electrostatically while its reflectance spectrum is measured with a spectrometer and a calibrated camera. The white light source is either a halogen lamp (for spectral studies) or a white LED synchronized to the membrane electrical control (for stroboscopic studies). c) At different voltages, the reflectance spectrum peak of the overall membrane shifts from 580~nm to 510~nm in a controlled and reproducible manner. Black lines are the theoretical spectral reflectance obtained from equation~(\ref{eqn:Eq4}).}
	\label{fgr:Fig1}
\end{figure}

\noindent
Circular cavities are etched through thermally-grown SiO$_2$ layers (300 to 1180~nm in depth depending on the sample) on silicon substrates by means of reactive-ion etching. DLG layers, fabricated by stacking two CVD single-layer graphene (SLG) layers, are transferred onto the patterned substrate using a semi-dry transfer technique. This results in an optical cavity with a movable absorbing membrane made out of CVD DLG (Figure~\ref{fgr:Fig1}a), and a fixed mirror formed by the underlying silicon surface. The silicon substrate also plays the role of back electrode for electrostatic actuation.
We use DLG membranes because of their higher yield~\cite{Cartamil-Bueno2017} and their larger absorption in the visible spectrum than that of SLG.

\noindent
The colorimetry setup used in this work consists of an optical microscope with K\"{o}hler illumination, a 20$\times$ apochromatic objective lens, and a halogen lamp as a multi-wavelength light source. Light reflected from the sample is split and guided both towards a calibrated consumer camera and a spectrometer. The spectrometer is configured to collect only the light from a circular area with the same size as the studied drums (see Methods and Supplementary Information Section 1). A series of color filters with wavelengths ranging from 450~nm to 650~nm (bandpass linewidth of 5-10~nm) are mounted on a motorized computer-controlled wheel, and placed in front of the camera imager as shown schematically in Figure~\ref{fgr:Fig1}b.

\noindent
Electrostatic actuation of the graphene drums is achieved by using a computer controlled DC source connected to the suspended graphene electrode, while keeping the silicon substrate grounded. For these measurements a sample with 1180~nm oxide thickness was chosen. The capacitance of the graphene/SiO$_2$/Si structure (area of $\sim$1~cm$^2$) was measured to be 0.6~nF.
Figure~\ref{fgr:Fig1}c shows the shift in the spectrum upon the application of a voltage. As the voltage is ramped up (see Methods), the overall color of the drumhead changes, going from orange ($\lambda=580$~nm) at 0~V to green ($\lambda=550$~nm at 12.5~V and $\lambda=510$~nm at 20~V).

\noindent
The spectral response of the graphene drums allows us to extract the deflection of the drum, which can be used to relate the electromechanical and optomechanical performance of the pixel.
By applying a force equilibrium condition between the electrostatic force, the linear elastic force (stiffness $k_1$), the cubic elastic force (stiffness $k_3$) and the hydrostatic force due to the pressure of trapped gas inside the cavity, we develop a 1-degree-of-freedom graphene membrane electromechanical model where the bending rigidity is neglected (detailed mathematical derivations in Supplementary Information Sections~2-4).
Then, we introduce the optical reflectance~\cite{Blake2007} by considering the suspended drum as an absorbing layer placed in front of the silicon back mirror while neglecting optical cavity effects.
We average the reflectance over the entire drum to obtain the response as perceived by the spectrometer (Supplementary Information Section~5). This "drum reflectance" can be expressed only in terms of the center deflection as
\begin{eqnarray}
R_{avg}(\lambda) = A(\lambda) + \frac{B(\lambda)}{\pi \bar{X}} \left[ \sin(\phi) - \sin(\phi - \pi \bar{X}) \right],
\label{eqn:Eq4}
\end{eqnarray}

\noindent
where $\lambda$ is the wavelength, $A(\lambda)$ and $B(\lambda)$ are wavelength dependent constants, $\phi = \frac{4 \pi g}{\lambda} + \phi'$ is a phase shift induced by the optical travel path and by the graphene, $\hat{X}\left(r \right) = \frac{4 g \bar{\delta}(r)}{\lambda}$ with $\bar{\delta}$ being the normalized deflection at the center of the drum (i.e. $\bar{\delta}=\delta_{center} /g$) and $g$, the cavity depth. This model is applied to our devices and shown in Figure~\ref{fgr:Fig1}c (solid lines).

\begin{figure}[h]
	\centering
	\graphicspath{{Figures/}}
	\includegraphics[width=177.8mm]{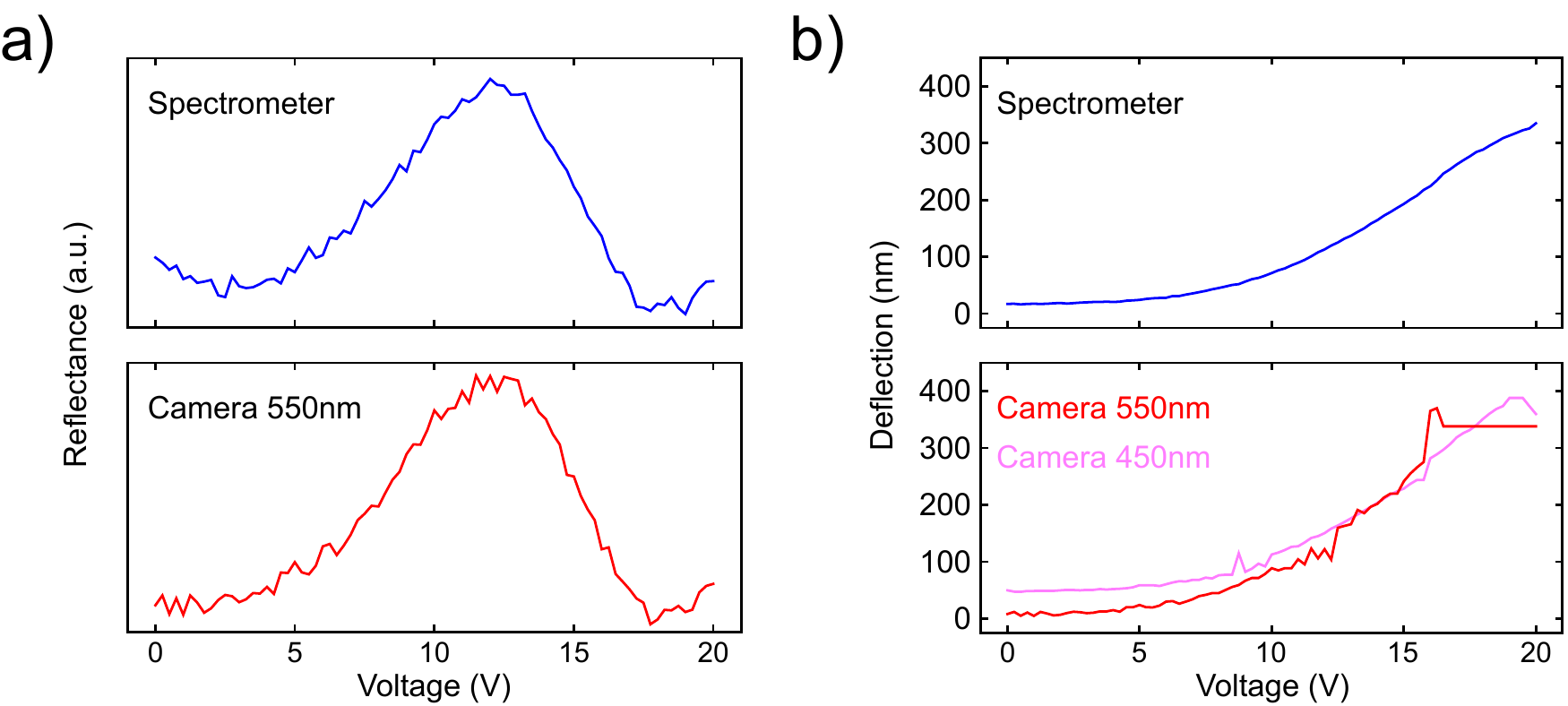}
	\caption{Conversion from reflectance-voltage curves to deflection-voltage curves for $\lambda=550$~nm. a) The optical response in that selected wavelength from the spectrometer measurements (top panel) and color-filtered camera images (bottom panel) show the same trend. b) By fitting the reflectance curves with equation~(\ref{eqn:Eq4}), we obtain the corresponding deflections as a function of applied voltage, which agree well for both types of measurements. A maximum deflection of 350-400~nm is obtained for this membrane (same device as in Figure~\ref{fgr:Fig1}).}
	\label{fgr:Fig2}
\end{figure}

\noindent
The electro-optic response of a suspended 20~$\mu$m in diameter graphene drum, i.e. the drum averaged reflectance as a function of voltage, is shown in Figure~\ref{fgr:Fig2}a as obtained from the camera and spectrometer for the 550~nm wavelength.
Using equation~(\ref{eqn:Eq4}), we fit a value of deflection, and plot the deflection vs. voltage in Figure~\ref{fgr:Fig2}b. The deflection curves from both camera and spectrometer agree well, although the camera fit is done on collected data from a narrow wavelength band (corresponding to that of the filter) while the spectrometer fit is done on a 475-700~nm spectral band. It is worth noting that the fits are insensitive to small deflections and thus to low actuation voltages.

\begin{figure}[h]
	\centering
	\graphicspath{{Figures/}}
	\includegraphics[width=177.8mm]{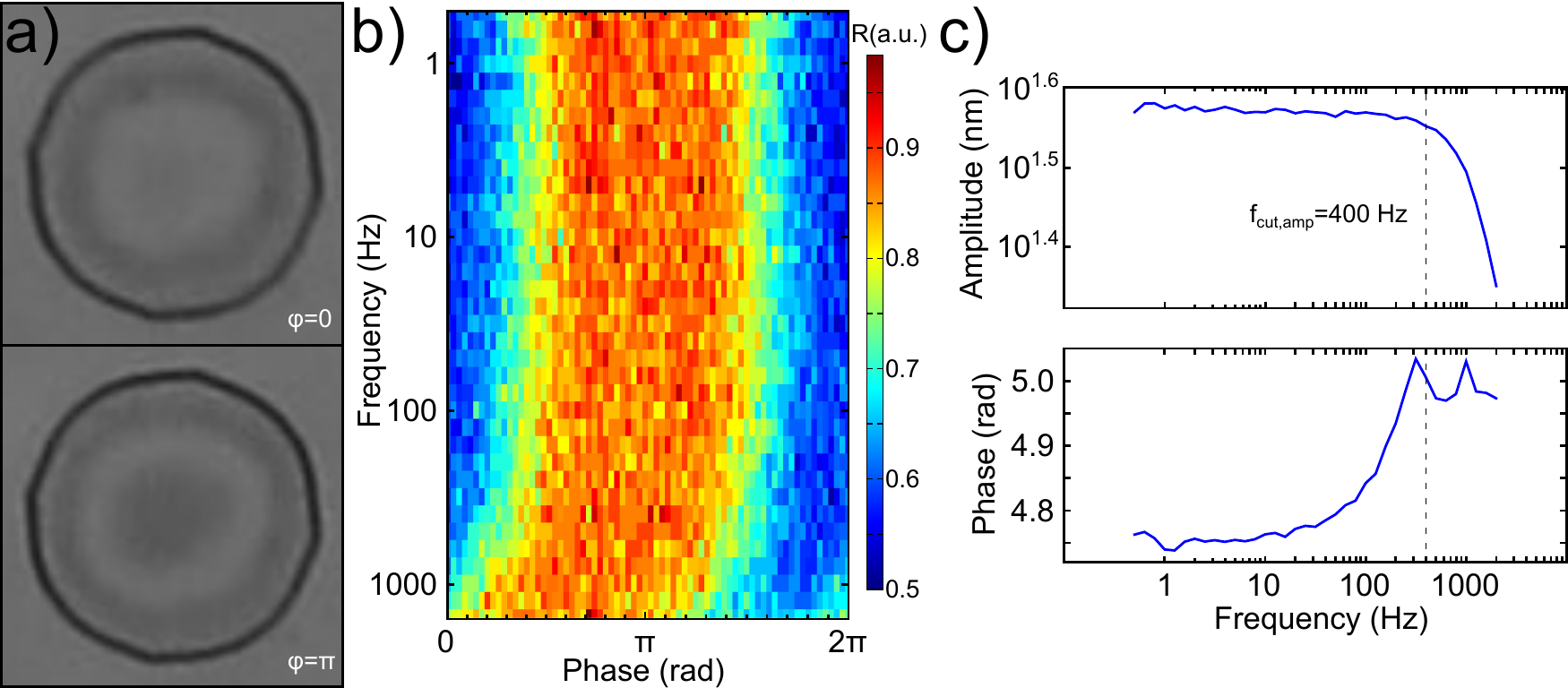}
	\caption{Stroboscopic measurement. a) Optical microscope image of a 15~$\mu$m in diameter drum at two opposite phases of vibration ($\phi=0$ top, $\phi=\pi$ bottom) while actuated at 1~kHz. b) Colormap of the reflectance at the center of the same drum as a function of frequency and phase. A phase delay intrinsic to the signal amplifier is observed. c) Bode plots of amplitude and phase of the drum, showing that the response is flat up to 400~Hz (cutoff frequency of amplifier).}
	\label{fgr:Fig3}
\end{figure}

\noindent
Given the gas impermeability of graphene~\cite{Bunch2008}, these drums should be hermetic, although they could permeate if few defects are present in the membrane~\cite{Cartamil-Bueno2016}.
In the previous measurement, a 30~s settling time was used to make sure that no gas permeation effects would interfere with the electro-optic response of the drum, but gas compression effects could alter the position of the membranes when actuating them at sufficiently higher frequencies.
Therefore, we investigate the possibility of a squeeze-film like response by fast stroboscopic measurements that are enabled by substituting the halogen lamp with a white LED, which is powered by short duration pulses that are synchronized to a sinusoidal actuation voltage: $V_{act} = (17.5 + 2.5 \sin(\omega_{act} t + \phi_{act}))$~V (see Methods). By sweeping the phase difference between the illuminating pulse and the actuation signal we are able to reconstruct the time-domain response of the structure. An example is shown in Figure~\ref{fgr:Fig3}a where we see two snapshots of a 15~$\mu$m drumhead at opposing phases in the actuation cycle.

We stroboscopically measure 5 drumheads of 15~$\mu$m in diameter with actuation frequencies ranging from 0.5 to 2000~Hz. The measured structures do not exhibit a first-order low-pass filter response indicative of squeeze-film effect for frequencies up to 400~Hz, beyond which the cut-off frequency of the capacitively loaded amplifier is reached. The frequency response of the drum's center reflectance is shown in Figure~\ref{fgr:Fig3}b. The absence of a first-order response indicates that the permeation time constants are below or above the measurement range depending on the drumhead in question.
A Bode plot of the displacement amplitude and phase are shown in Figure~\ref{fgr:Fig3}c, demonstrating no special feature until the amplifier cut-off around 400~Hz.

\begin{figure}[h]
	\centering
	\graphicspath{{Figures/}}
	\includegraphics[width=177.8mm]{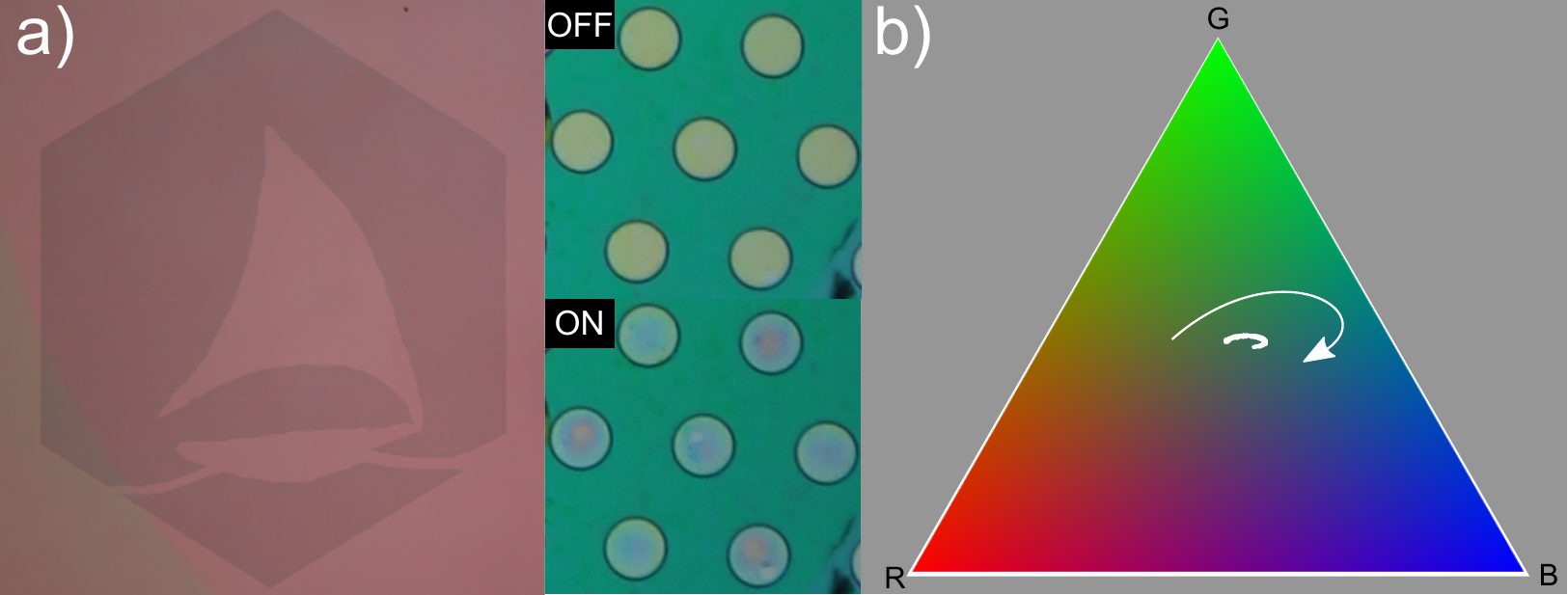}
	\caption{Graphene Interferometric MOdulator Display (GIMOD) prototype showing the logo of the Graphene Flagship. The graphene mechanical devices are used as continuous-spectrum pixels of 5~$\mu$m in diameter, hence resulting in a reflective-type display of 2500~pixels per inch (ppi), equivalent to 12K+ resolution for a display of 5$''$. When the GIMOD prototype is OFF, its pixels show a yellow color; when it is ON (30~V), they are in a blue state. b) Gamut pixel trajectory of the average reflectance of a GIMOD prototype with pixels of 5~$\mu$m in diameter. The change of pixel color as a function of voltage, whose direction is pointed with an arrow, is displayed in the sRGB color triangle (subset of x,y chromaticity space) based on CIE1931 colorimetry.}
	\label{fgr:Fig4}
\end{figure}

\noindent
The large electro-optic modulation and absence of mechanical delays at frequencies up to 400~Hz make these devices interesting to enable display rates beyond the flicker fusion threshold that require color reproducibility in the millisecond range for applications such as virtual or augmented reality (VR/AR)~\cite{davis2015}.
Figure~\ref{fgr:Fig4}a shows the substrate of a $\sim$2500~ppi Graphene Interferometric MOdulator Display (GIMOD) prototype. The static image displaying the Graphene Flagship logo changes color upon the application of a sinusoidal signal (see Supplementary Video S10). All the mechanical pixels of 5~$\mu$m in diameter are addressed at once with the silicon substrate acting as a back gate. The pixel aperture -active to total area ratio- is $\sim$50\%, resulting in a reduction of contrast of about 1:2 due to the large separation between pixels (pixel pitch about double of the pixel size). The panels in the middle show a zoom-in of some pixels when they are actuated with 30~V (yellow state when OFF, top panel; blue state when ON, bottom panel).

By analyzing the RGB channels of one of the pixels, we obtain its average color gamut and its evolution upon applied voltage. Figure~\ref{fgr:Fig4}b shows the pixel trajectory over a standard RGB (sRGB) color map based on CIE1931 colorimetry~\cite{westland2012}, demonstrating the electro-optical modulation from yellow to blue of these pixels.

While electrostatic transduction is used extensively as an actuation mechanism for MEMS/NEMS devices, including graphene nanoelectromechanical structures~\cite{Bunch2007,Reserbat-Plantey2012,AbdelGhany2016}, in this work we investigate the far-from-resonance, quasi-static, large deformation regime of graphene membranes~\cite{Rokni2013,AbdelGhany2012}.
For such a regime of operation the hydrostatic pressure component near the zero-deflection position ($\bar{\delta} \sim 0$) should not be neglected in hermetic cavities as it introduces a significant hydrostatic stiffness (Supplementary Information Section~4). Only if the membrane is not sufficiently hermetic can this effect be neglected. Moreover, large deflections --even in the case of impermeable membranes-- cause the hydrostatic component to be overtaken by the nonlinear stiffness. However, it is possible to have structures with large diameters ($>5$~$\mu$m) and shallow cavities ($<300$~nm) whereby for a hermetic cavity the hydrostatic stiffness remains dominant over the nonlinear one for all deflections. In such cases both hydrostatic stiffness and gas permeation time can interfere with the proper functioning of the device.

It is interesting to note that amongst the five 15~$\mu$m in diameter membranes measured  stroboscopically, one drum shows larger static and dynamic deflections compared to the others (Supplementary Information Section~6). Whether this is due to a lower mechanical stiffness caused by micro-tears, to a higher membrane permittivity, or to a combination of both (they are not mutually exclusive), remains a question for further investigation.

The measured DC voltage-dependent spectrum is qualitatively similar to simulation results obtained from equation~(\ref{eqn:Eq4}) (Supplementary Information Section~5). As it can be also seen in Figure~\ref{fgr:Fig2}b, the membrane deflection and measured spectrum show little change for low voltages ($<5$~V). Since drums of 20~$\mu$m in diameter are rarely hermetic, this effect is more likely explained by limitations of the technique itself rather than to hydrostatic stiffness. In particular, the assumption of a parabolic profile although accurate for large deflections, is not necessarily valid for small deformations due to wrinkles~\cite{Nicholl2012}. This is a necessary tradeoff in order to fit accurate deflection values for SLG/DLG due to the need to perform a drum average fit rather than a pixel-by-pixel fit as done for thicker membranes~\cite{Northeast2017}.

Furthermore, it is interesting to note that according to equation~(\ref{eqn:Eq4}), the electro-optic response of a GIMOD pixel, when averaged over the drum area, is radius independent. Thus a change in the radius of the suspended drums does not result in any qualitative modification of the pixel behavior. In addition, the obtained RGB response of the pixels as shown in Figure~\ref{fgr:Fig4}, shows a limited path across the sRGB triangle. This is confirmed by simulations that predict a maximum change of 10\% for each of the RGB channels for a DLG membrane and for an optimal gap for a GIMOD of 600~nm (Supplementary Information Section~7). For a richer colorimetric response, thicker membranes may be required. Ultimately graphene mechanical pixels need to be optimized taking into consideration the yield of the fabrication process, the actuation voltage, and the colorimetric response.

\noindent
In summary, we report on CVD double-layer graphene electromechanical electro-optic modulators, the characterization of their spectral response, and their application as pixels in a Graphene Interferometric MOdulator Display (GIMOD).
In addition to observing the electro-optical modulation, we develop an electrostatic-optical model to describe their behavior. We use the reflectance-voltage plots to fit a deflection-voltage response and conclude that the device is not significantly hermetic. We further explored the permeation time constants by stroboscopically measuring 5 devices without finding the onset of squeeze-film effect due to the gas trapped in the cavity in the frequency range of 0.5 to 400~Hz.
The large color modulation and good frequency response of these graphene electro-optic modulators in ambient conditions enable the use of these mechanical devices as pixels with color reliability at high image refresh rates.
This application is demonstrated with a $0.5''$ prototype with 2500~ppi that would equate to a $5''$ display with 12K+ resolution.

Besides possibly eliminating the motion sickness in VR/AR applications, the demonstrated reflective-type (e-paper) pixel technology would greatly reduce power consumption as it is an Interferometric MOdulator display (IMOD) technology. The use of graphene allows the large deformation of small membranes without mechanical failure or hysteresis allowing the fabrication of diffraction-limited devices for ultra-high resolution displays. Moreover, by using different voltages to actuate the pixels, each device is able to generate natural colors in the full-spectrum and in a continuous manner. This eliminates the need for RGB subpixels and reduces the addressing bandwidth to only one channel, which enables reduction of power consumption.

\section*{Methods}
\noindent
\textbf{Colorimetry technique with optical spectrometer} The deflection as a function of voltage are obtained with the colorimetry technique with an Olympus BX60 microscope as previously reported~\cite{Cartamil-Bueno2016,Cartamil-Bueno2017}. Pictures are taken with a Canon EOS 600D and a 20$\times$ Olympus UMPlanFI objective lens (NA 0.46, LWD 3~mm). A TE-cooled linear CCD array spectrometer (B\&W Tek Glacier X) is attached to the optical microscope in such a way it collects the light reflected from an area slightly larger than the selected drum (about 595~$\mu$m$^2$) (see Supplementary Information). Spectral traces of the entire drum are recorded with an integration time of 20~ms (no averaging). For the electrostatic pulling, we use a power supply (Rigol DP832) to ramp up the voltage from 0 to 20~V with steps of 0.25~V. The voltage is left constant for a period of 30~s after each step and before acquiring the data.

\noindent
\textbf{Stroboscopic technique} The halogen lamp from the microscope is substituted by a white LED for stroboscopic illumination. A dual channel generator (Keysight Technologies 33512B Waveform Generator) controls the LED by outputting a narrow ($\frac{1}{72}$ of the actuation period) 5~V pulse to the LED. At the same time this illumination pulse is phase-locked to an actuation sine wave that drives the graphene membrane electrostatically.

\section*{Data Availability}
The authors declare that the data supporting the findings of this study are available within the paper and its supplementary information files.

\begin{addendum}
	\item The research leading to these results has received funding from the European Union's Horizon 2020 research and innovation programme under grant agreement No 649953 (Graphene Flagship).
	\item[Competing Interests] The authors declare that they have no competing financial interests.
	\item[Correspondence] Correspondence and requests for materials should be addressed to Santiago J. Cartamil-Bueno~(email: cartamil@amo.de).
	\item[Supplementary Information] Supplementary information accompanies the manuscript on the \textit{Nature Nanotechnology} website http://www.nature.com/naturenanotech
	\item[Contributions] Devices were fabricated by S.J.C.-B. and D.D. Experiments were designed and performed by S.J.C.-B. and S.H. Materials were designed and synthesized by A.C. and A.Z. Data was analyzed by S.J.C.-B. and S.H., and all authors contributed to the discussion. S.J.C.-B. and S.H. wrote the manuscript, with contributions from all authors.
\end{addendum}

\section*{References}
\nointerlineskip
\noindent

\begin{thebibliography}{10}
	\expandafter\ifx\csname url\endcsname\relax
	\def\url#1{\texttt{#1}}\fi
	\expandafter\ifx\csname urlprefix\endcsname\relax\def\urlprefix{URL }\fi
	\providecommand{\bibinfo}[2]{#2}
	\providecommand{\eprint}[2][]{\url{#2}}
	
	\bibitem{Bonaccorso2010}
	\bibinfo{author}{Bonaccorso, F.}, \bibinfo{author}{Sun, Z.}, \bibinfo{author}{Hasan, T.} \& \bibinfo{author}{Ferrari, A.}
	\newblock \bibinfo{title}{Graphene photonics and optoelectronics}.
	\newblock \emph{\bibinfo{journal}{Nature Photonics}}
	\textbf{\bibinfo{volume}{4}}, \bibinfo{pages}{611--622}
	(\bibinfo{year}{2010}).

	\bibitem{Yu2015}
	\bibinfo{author}{Yu, R.}, \bibinfo{author}{Pruneri, V.} \& \bibinfo{author}{García de Abajo, F.J.}
	\newblock \emph{\bibinfo{title}{Resonant visible light modulation with graphene}}.
	\newblock \emph{\bibinfo{journal}{ACS Photonics}}
	\textbf{\bibinfo{volume}{2}} (\bibinfo{year}{2015}).
	
	\bibitem{Sun2016}
	\bibinfo{author}{Sun, Z.}, \bibinfo{author}{Martinez, A.} \& 	\bibinfo{author}{Wang, F.}
	\newblock \bibinfo{title}{Optical modulators with 2d layered materials}.
	\newblock \emph{\bibinfo{journal}{Nature Photonics}}
	\textbf{\bibinfo{volume}{10}}, \bibinfo{pages}{227--238}
	(\bibinfo{year}{2016}).

	\bibitem{Ferrari2014}
	\bibinfo{author}{Ferrari, A.~C.} \emph{et~al.}
	\newblock \bibinfo{title}{{Science and technology roadmap for graphene, related two-dimensional crystals, and hybrid systems}}.
	\newblock \emph{\bibinfo{journal}{Nanoscale}} \textbf{\bibinfo{volume}{7}},
	\bibinfo{pages}{4598--4810} (\bibinfo{year}{2014}).

	\bibitem{Chan2017}
	\bibinfo{author}{Chan, E.~K.} \emph{et~al.}
	\newblock \bibinfo{title}{Continuous color reflective display fabricated in integrated mems-and-tft-on-glass process}.
	\newblock \emph{\bibinfo{journal}{Journal of Microelectromechanical Systems}}
	\textbf{\bibinfo{volume}{26}}, \bibinfo{pages}{143--157}
	(\bibinfo{year}{2017}).

	\bibitem{Cartamil-Bueno2016}
	\bibinfo{author}{Cartamil-Bueno, S.~J.} \emph{et~al.}
	\newblock \bibinfo{title}{{Colorimetry Technique for Scalable Characterization of Suspended Graphene}}.
	\newblock \emph{\bibinfo{journal}{Nano Letters}} \textbf{\bibinfo{volume}{16}},
	\bibinfo{pages}{6792--6796} (\bibinfo{year}{2016}).

	\bibitem{Cartamil-Bueno2017}
	\bibinfo{author}{Cartamil-Bueno, S.~J.} \emph{et~al.}
	\newblock \bibinfo{title}{{Very large scale characterization of graphene mechanical devices using a colorimetry technique}}.
	\newblock \emph{\bibinfo{journal}{Nanoscale}} \textbf{\bibinfo{volume}{9}},
	\bibinfo{pages}{7559--7564} (\bibinfo{year}{2017}).
	
	\bibitem{Blake2007}
	\bibinfo{author}{Blake, P.} \& \bibinfo{author}{Hill, E.}
	\newblock \bibinfo{title}{{Making graphene visible}}.
	\newblock \emph{\bibinfo{journal}{Applied Physics Letters}}
	\textbf{\bibinfo{volume}{063124}}, \bibinfo{pages}{3} (\bibinfo{year}{2007}).
	
	\bibitem{Bunch2008}
	\bibinfo{author}{Bunch, J.~S.} \emph{et~al.}
	\newblock \bibinfo{title}{{Impermeable atomic membranes from graphene sheets.}}
	\newblock \emph{\bibinfo{journal}{Nano Letters}}
	\textbf{\bibinfo{volume}{8}}, \bibinfo{pages}{2458--2462}
	(\bibinfo{year}{2008}).
	
	\bibitem{davis2015}
	\bibinfo{author}{Davis, J.}, \bibinfo{author}{Hsieh, Y.-H.} \&
	\bibinfo{author}{Lee, H.-C.}
	\newblock \bibinfo{title}{Humans perceive flicker artifacts at 500 Hz}.
	\newblock \emph{\bibinfo{journal}{Scientific Reports}}
	\textbf{\bibinfo{volume}{5}} (\bibinfo{year}{2015}).
	
	\bibitem{westland2012}
	\bibinfo{author}{Westland, S.}, \bibinfo{author}{Ripamonti, C.} \&
	\bibinfo{author}{Cheung, V.}
	\newblock \emph{\bibinfo{title}{Computational colour science using MATLAB}}
	(\bibinfo{publisher}{John Wiley \& Sons}, \bibinfo{year}{2012}).
	
	\bibitem{Bunch2007}
	\bibinfo{author}{Bunch, J.~S.} \emph{et~al.}
	\newblock \bibinfo{title}{{Electromechanical resonators from graphene sheets.}}
	\newblock \emph{\bibinfo{journal}{Science}}
	\textbf{\bibinfo{volume}{315}}, \bibinfo{pages}{490--493}
	(\bibinfo{year}{2007}).
	
	\bibitem{AbdelGhany2016}
	\bibinfo{author}{AbdelGhany, M.} \emph{et~al.}
	\newblock \bibinfo{title}{Suspended graphene variable capacitor}.
	\newblock \emph{\bibinfo{journal}{2D Materials}} \textbf{\bibinfo{volume}{3}},
	\bibinfo{pages}{041005} (\bibinfo{year}{2016}).

	\bibitem{Reserbat-Plantey2012}
	\bibinfo{author}{Reserbat-Plantey, A.}, \bibinfo{author}{Marty, L.},
	\bibinfo{author}{Arcizet, O.}, \bibinfo{author}{Bendiab, N.} \& \bibinfo{author}{Bouchiat, V.}
	\newblock \bibinfo{title}{{A local optical probe for measuring motion and stress in a nanoelectromechanical system.}}
	\newblock \emph{\bibinfo{journal}{Nature Nanotechnology}}
	\textbf{\bibinfo{volume}{7}}, \bibinfo{pages}{151--5} (\bibinfo{year}{2012}).

	\bibitem{Rokni2013}
	\bibinfo{author}{Rokni, H.} \& \bibinfo{author}{Lu, W.}
	\newblock \bibinfo{title}{Effect of graphene layers on static pull-in behavior of bilayer graphene/substrate electrostatic microactuators}.
	\newblock \emph{\bibinfo{journal}{Journal of Microelectromechanical Systems}}
	\textbf{\bibinfo{volume}{22}}, \bibinfo{pages}{553--559}
	(\bibinfo{year}{2013}).
	
	\bibitem{AbdelGhany2012}
	\bibinfo{author}{AbdelGhany, M.}, \bibinfo{author}{Ledwosinska, E.} \&
	\bibinfo{author}{Szkopek, T.}
	\newblock \bibinfo{title}{Theory of the suspended graphene varactor}.
	\newblock \emph{\bibinfo{journal}{Applied Physics Letters}}
	\textbf{\bibinfo{volume}{101}}, \bibinfo{pages}{153102}
	(\bibinfo{year}{2012}).
	
	\bibitem{Boddeti2013}
	\bibinfo{author}{Boddeti, N.~G.} \emph{et~al.}
	\newblock \bibinfo{title}{{Mechanics of Adhered, Pressurized Graphene Blisters}}.
	\newblock \emph{\bibinfo{journal}{Journal of Applied Mechanics}}
	\textbf{\bibinfo{volume}{80}}, \bibinfo{pages}{40909} (\bibinfo{year}{2013}).
	
	\bibitem{Li2013}
	\bibinfo{author}{Li, Z.} \emph{et~al.}
	\newblock \bibinfo{title}{Resonant frequency analysis on an electrostatically actuated microplate under uniform hydrostatic pressure}.
	\newblock \emph{\bibinfo{journal}{Journal of Physics D: Applied Physics}}
	\textbf{\bibinfo{volume}{46}}, \bibinfo{pages}{195108}
	(\bibinfo{year}{2013}).
		
	\bibitem{Nicholl2012}
	\bibinfo{author}{Nicholl, R.J.T.}, \bibinfo{author}{Lavrik, N.V.}, \bibinfo{author}{Vlassiouk, I.}
	\bibinfo{author}{Srijanto, B.R.} \& \bibinfo{author}{Bolotin, K. I.}
	\newblock \emph{\bibinfo{title}{Hidden area and mechanical nonlinearities in freestanding graphene}}.
	\newblock \emph{\bibinfo{journal}{Physical Review Letters}}
	\textbf{\bibinfo{volume}{118}} (\bibinfo{year}{2017}).

	\bibitem{Northeast2017}
	\bibinfo{author}{Northeast, D.B.}, \& \bibinfo{author}{Knobel, R.G.}
	\newblock \emph{\bibinfo{title}{Fabrication and optical characterization of suspended 2D membranes for electromechanics}}.
	\newblock \emph{\bibinfo{journal}{Preprint at https://arxiv.org/abs/1710.11320}}
	\textbf{\bibinfo{volume}{}} (\bibinfo{year}{2017}).
\end{thebibliography}

\end{document}


\maketitle

\begin{affiliations}
	\item Advanced Microelectronic Center Aachen (AMICA), AMO GmbH, Otto-Blumenthal-Str. 25, 52074 Aachen, Germany
	\item Kavli Institute of Nanoscience, Delft University of Technology, Lorentzweg 1, 2628CJ, Delft, The Netherlands
	\item Graphenea SA, 20018 Donostia-San Sebasti\'an, Spain\\
$^*$Corresponding authors: cartamil@amo.de., Houri.Samer@lab.ntt.co.jp.
\end{affiliations}
\nointerlineskip
\noindent{\it Keywords}: CVD graphene, electro-optic modulators, MEMS, display, GIMOD
	
\setcounter{equation}{0}
\setcounter{figure}{0}
\setcounter{table}{0}
\renewcommand\theequation{S-\arabic{equation}}
\renewcommand\thefigure{S-\arabic{figure}}
\renewcommand\thetable{S-\arabic{table}}

\section*{Supplementary Information}
\nointerlineskip
Supplementary Information Outline:\\
\noindent
1. Area measured by spectrometer\\
2. Capacitance of the deflected circular membrane\\
3. Electrostatic force on a circular graphene membrane\\
4. Force equilibrium equation\\
5. Optical response of deflected drums\\
6. Stroboscopic measurement of graphene drums/pixels\\
7. Simulating the RGB response of GIMOD pixels\\

\nointerlineskip
\section*{1. Area measured by spectrometer}
\begin{figure}[h]
	\centering
	\graphicspath{{Figures/Supplementary_Info/}}
	\includegraphics[width=177.8mm]{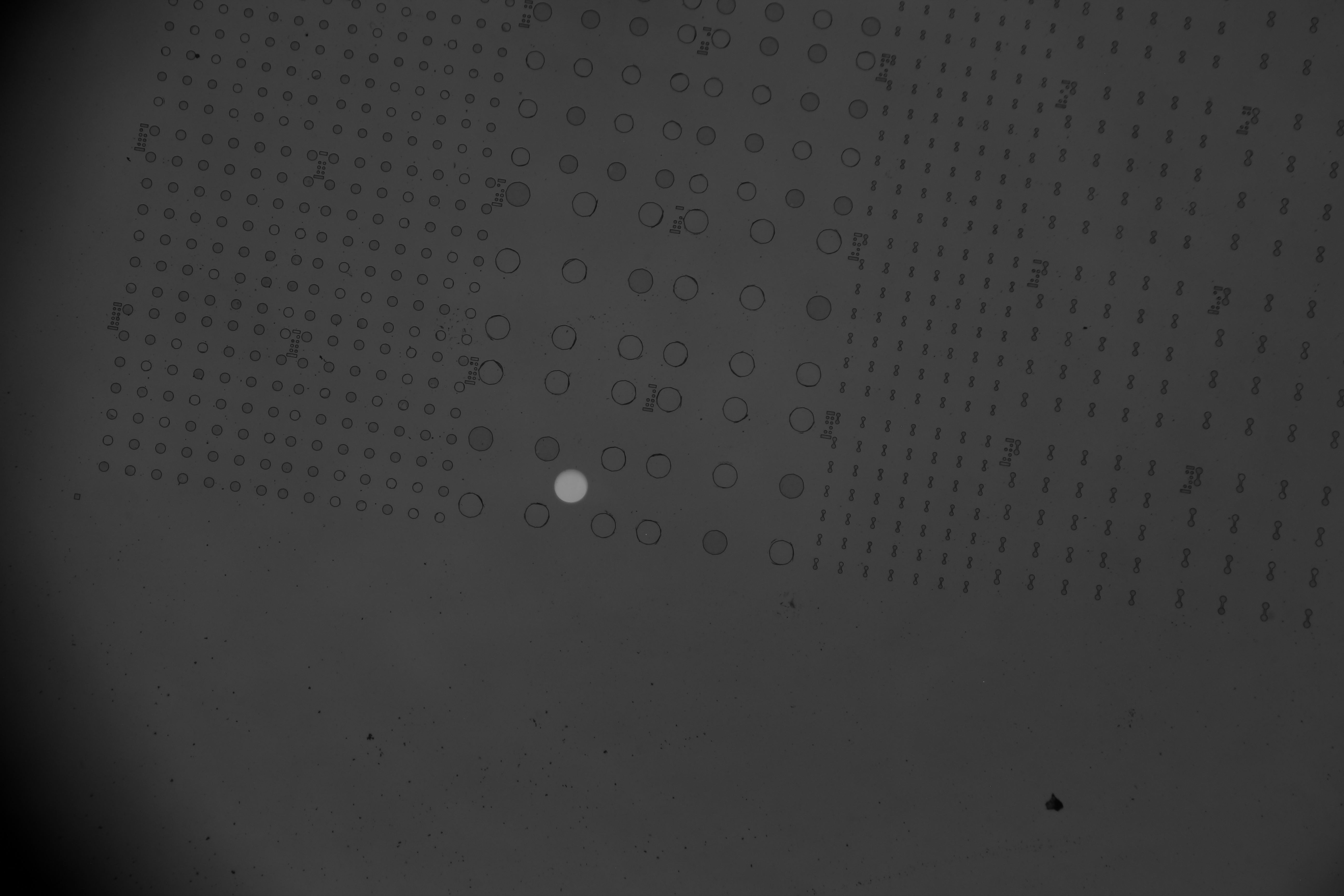}
	\caption{Optical microscope image showing the light collection area of the spectrometer. Here the spectrometer was replaced by a light source so as to project a spot on the sample.}
	\label{fgr:FigS1}
\end{figure}

\nointerlineskip
For the spectrometer measurements, we used a focused fiber to collect the reflected light from the graphene drumheads. In order to align the fiber so it points and covers the desired device, we first connect the fiber to a light source so we can observe a bright spot on top of the sample. Figure~\ref{fgr:FigS1} shows the light spot of 27.5~$\mu$m in diameter located away from the devices, covering an area of graphene on SiO$_2$.

\section*{2. Capacitance of the deflected circular membrane}
\nointerlineskip
We apply a 1 degree of freedom graphene membrane model that neglects bending rigidity, and describes the drums' deflection by an axisymmetric parabolic profile given by $\delta(r) = \delta_c\left(1-(\frac{r}{a})^2\right)$, where $\delta_c$ is the center deflection, $a$ is the radius of the drum, and $r$ is the radial distance away from the drum's center. The capacitance of the circular membrane is now given by:
\begin{equation}
\ C = \int_{0}^{2\pi} {\int_{0}^{a}{\frac{\epsilon_{0}rdrd\theta}{g_0-\delta(r)}}}\\
\label{eqn:Eq1}
\end{equation}

\noindent
We define $\bar{\delta} = \frac{\delta_c}{g_0}$, $Y = \left(1-\frac{r^2}{a^2}\right)$, and $C_0 = \frac{\epsilon_0\pi a^2}{g_0}$, and re-express equation~\ref{eqn:Eq1} as:\\
\begin{equation}
\begin{split}
\ C = -C_0\int_{1}^{0} {\frac{dY}{1-\bar{\delta}Y}} = \frac{C_0}{\bar{\delta}}\int_{1}^{0} {\frac{d(1-\bar{\delta}Y)}{1-\bar{\delta}Y}}\\
\ C = -C_0\frac{ln(1-\bar{\delta})}{\bar{\delta}}
\end{split}
\label{eqn:Eq2}
\end{equation}

\section*{3. Electrostatic force on a circular graphene membrane}
\nointerlineskip
We obtain the electrostatic force under the effect of an applied voltage $V$ by taking the spatial derivative of the electrostatic potential energy, thus:
\begin{equation}
\begin{split}
\ F_{Elec}(\delta_c,V) = -\frac{d}{d\delta_c} \frac{1}{2}CV^2 = -\frac{1}{g_0}\frac{d}{d\bar{\delta}} \frac{1}{2}CV^2 \\
\ F_{Elec} = -\frac{C_0 V^2}{2 g_0}\left(\frac{\bar{\delta}-(\bar{\delta}-1)ln(1-\bar{\delta})}{(\bar{\delta}-1)\bar{\delta}^2} \right)\\
\end{split}
\label{eqn:Eq3}
\end{equation}

%

\section*{4. Force equilibrium equation}
\nointerlineskip
The deflection of the graphene membrane, if the perfect hermeticity assumption is to be maintained, requires that the force balance equation accounts for the hydrostatic pressure that develops from the compression of the gas trapped within the cavity and cannot escape it.
To calculate this last effect, we assume that the parabolic profile of membrane deflection applies and that the compression is isothermal, thus applying the perfect gas law the gas pressure within the cavity reads:
\begin{equation}
\ P = \frac{Ag_0P_0}{V} = \frac{Ag_0P_0}{A(g_0-\frac{\delta_c}{2})} = \frac{P_0}{1-\frac{\bar{\delta}}{2}}, \\
\label{eqn:Eq4}
\end{equation}

\noindent
where, $A$ is the area of the circular membrane, $P_0$ is the initial (ambient) pressure, and the factor $\frac{1}{2}$ in the denominator on the rightmost hand side term comes from integrating the area under the parabolic profile [S-1].
Thus the hydrostatic force acting on the membrane due to the change in the cavity pressure can be written as:
\begin{equation}
\begin{split}
\ F_{Hydro} = A\Delta P = A(P-P_0) = AP_0\frac{\bar{\delta}/2}{1-\bar{\delta}/2}\\
\label{eqn:Eq5}
\end{split}
\end{equation}
\noindent
Note that equation~\ref{eqn:Eq5} applies in case of both positive and negative deflection, i.e. positive and negative differential pressure.\\
Combining all the terms we obtain the following force equilibrium equation:
\begin{equation}
\begin{split}
\ k_1\delta_c + k_3\delta_c^3 + F_{Hydro} = F_{Elec}\\
\ => k_1 g_0\bar{\delta} + k_3g_0^3\bar{\delta}^3 + AP_0\frac{\bar{\delta}/2}{1-\bar{\delta}/2} = -\frac{C_0V^2}{2g_0}\left(\frac{\bar{\delta}-(\bar{\delta}-1)ln(1-\bar{\delta})}{(\bar{\delta}-1)\bar{\delta}^2} \right)\\
\end{split}
\label{eqn:Eq6}
\end{equation}

\begin{figure}[h]
	\centering
	\graphicspath{{Figures/Supplementary_Info/}}
	\includegraphics[width=84.6mm]{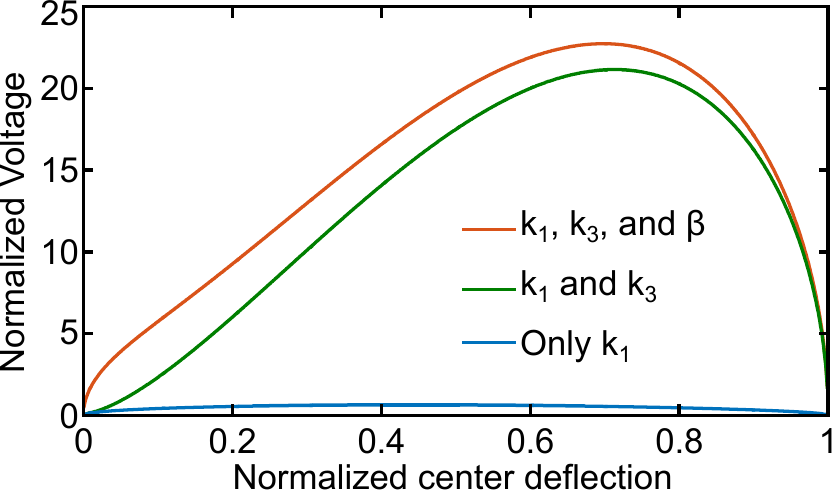}
	\caption{The voltage-deflection curves for pure linear stiffness (blue), combined linear and cubic stiffness (green), and combined linear, cubic and hydrostatic effects (red), as obtained for $k_3=1000 k_1$, and $\beta=100 k_1$.}
	\label{fgr:FigS2}
\end{figure}

\noindent
Equation~\ref{eqn:Eq6} can be rewritten in the following non-dimensional form:
\begin{equation}
\bar{\delta} + \bar{k_3}\bar{\delta}^3 + \beta\frac{\bar{\delta}/2}{1-\bar{\delta}/2} = -\frac{\bar{V}^2}{2}\left(\frac{\bar{\delta}-(\bar{\delta}-1)ln(1-\bar{\delta})}{(\bar{\delta}-1)\bar{\delta}^2} \right),\\
\label{eqn:Eq7}
\end{equation}
\noindent
where the non-dimensional parameters are given as $\bar{k_3} = k_3g_0^2/k_1$, $\beta = AP_0/k_1 g_0 $, and $\bar{V}^2 = C_0V^2/k_1 g_0$.\\
Typical values of $\bar{k_3}$ range between $10^{-1}-10^4$, of $\beta$ between $1-100$, and $\bar{V}^2$ between $(10^{-3}-1)V^2$.
In Figure~\ref{fgr:FigS2} the voltage displacement curve is plotted according to equation~\ref{eqn:Eq7} to show the impact of nonlinear stiffness and gas compression on the mechanical susceptibility of the device.
%

\section*{5. Optical response of deflected drums}
\nointerlineskip
The optical reflectivity model used neglects any cavity effects and treats the suspended drum as an absorbing layer placed in front of the silicon back mirror. This simplification is valid for thin graphene membranes, less than 5 layers [S-2], and leads to the drum reflectance
\begin{eqnarray}
R(r,\lambda) = A(\lambda) + B(\lambda) \cos (\phi-\pi \bar{X}(r)) \textnormal{, \,\,\,\,\,\,\,\,\, for \,\,} \bar{X}(r),
\label{eqn:Eq3n}
\end{eqnarray}
\noindent
where $\lambda$ is the wavelength, $A(\lambda)$ and $B(\lambda)$ are wavelength dependent constants, $\phi = \frac{4 \pi g}{\lambda} + \phi'$ is a phase shift induced by the optical travel path and by the graphene, $\hat{X}\left(r \right) = \frac{4g\bar{\delta}(r)}{\lambda}$ and $\delta(r) = \delta_c\left(1-(\frac{r}{a})^2\right)$. Taking nominal values for the optical properties of graphene, we can neglect $\phi'$ as it is negligibly small compared to the travel path term.
By integrating the reflectance over the drum's area, we obtain:
\begin{equation}
\begin{split}
R_{drum}\left(\lambda \right) = \int_{0}^{2\pi} {\int_{0}^{a}{R\left(r,\lambda \right) rdrd\theta}}\\
\ => R_{drum}\left(\lambda \right) = 2\pi \left[\frac{-a^2}{2}{\int_{1}^{0}{A(\lambda) + B(\lambda)cos\left(\phi - \frac{4\pi\delta_c}{\lambda}Y \right)dY}} \right]\\
\label{eqn:Eq9}
\end{split}
\end{equation}
\noindent
where $Y = \left(1-\frac{r^2}{a^2}\right)$.
Dividing equation~\ref{eqn:Eq9} by $\pi a^2$ to obtain the drum's average reflectivity, and integrating we arrive at:
\begin{equation}
R_{avg}\left(\lambda \right) =  A(\lambda) + B(\lambda)\frac{\lambda}{4\pi\delta_c}\left[ sin(\phi) - sin\left(\phi - \pi \frac{4\delta_c}{\lambda} \right) \right]\\
\label{eqn:Eq10}
\end{equation}

\begin{figure}[H]
	\centering
	\graphicspath{{Figures/Supplementary_Info/}}
	\includegraphics[width=81mm]{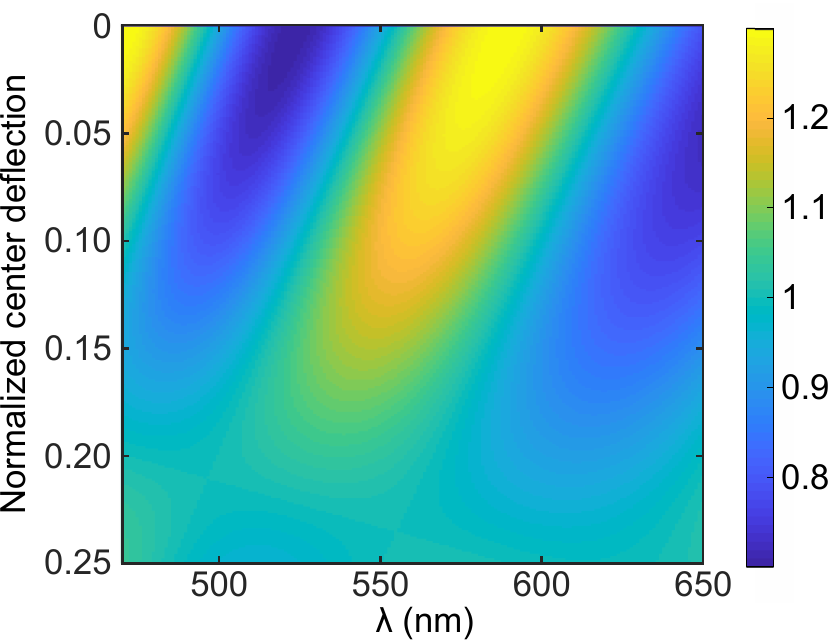}
	\caption{Simulation of the normalized drum averaged reflectance based on equation~\ref{eqn:Eq10}, using values for $A(\lambda)$ and $B(\lambda)$ from the fitted data.}
	\label{fgr:FigS3}
\end{figure}

\begin{figure}[H]
	\centering
	\graphicspath{{Figures/Supplementary_Info/}}
	\includegraphics[width=81mm]{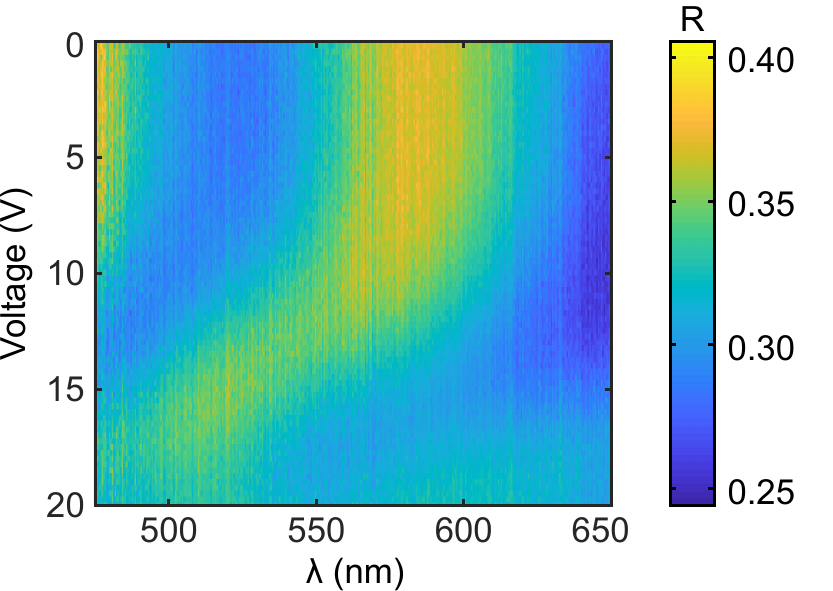}
	\caption{Measured drum averaged reflection as obtained from spectrometer data, shown as a function of electrostatic actuation voltage.}
	\label{fgr:FigS4}
\end{figure}

\noindent
Figure~\ref{fgr:FigS3} shows the simulated spectral response, as obtained from equation~\ref{eqn:Eq10}, of a graphene drum as a function of center deflection for a gap of 1140 nm. Figure~\ref{fgr:FigS4} shows the experimentally obtained spectral response as a function of voltage for the same gap.

\section*{6. Stroboscopic measurement of graphene drums/pixels}
\begin{figure}[H]
	\centering
	\graphicspath{{Figures/Supplementary_Info/}}
	\includegraphics[width=80mm]{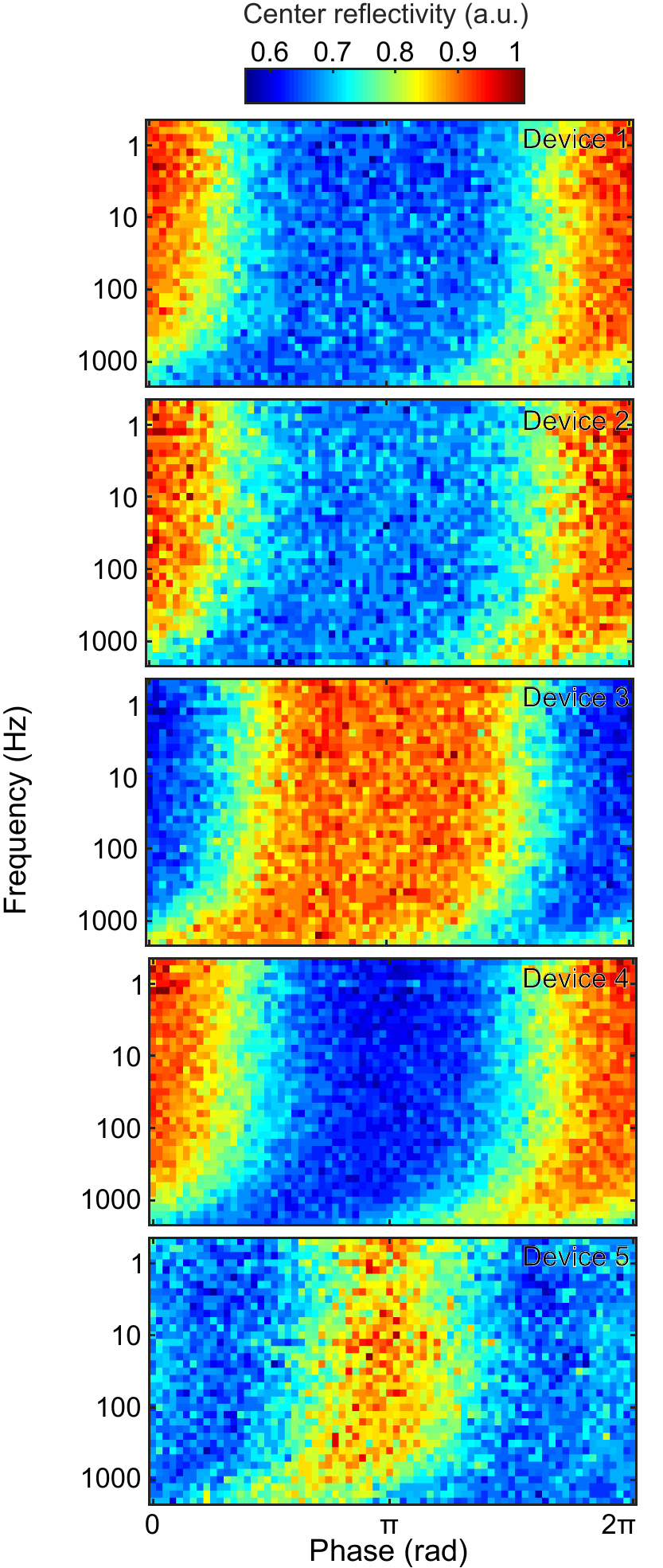}
	\caption{Experimentally obtained normalized reflectance at the center of all drums as a function of frequency and phase for all the drums measured stroboscopically. The same phase delay (from the signal amplifier) is observed as in Figure~3b from the main text.}
	\label{fgr:FigS5}
\end{figure}

\begin{figure}[H]
	\centering
	\graphicspath{{Figures/Supplementary_Info/}}
	\includegraphics[width=84.6mm]{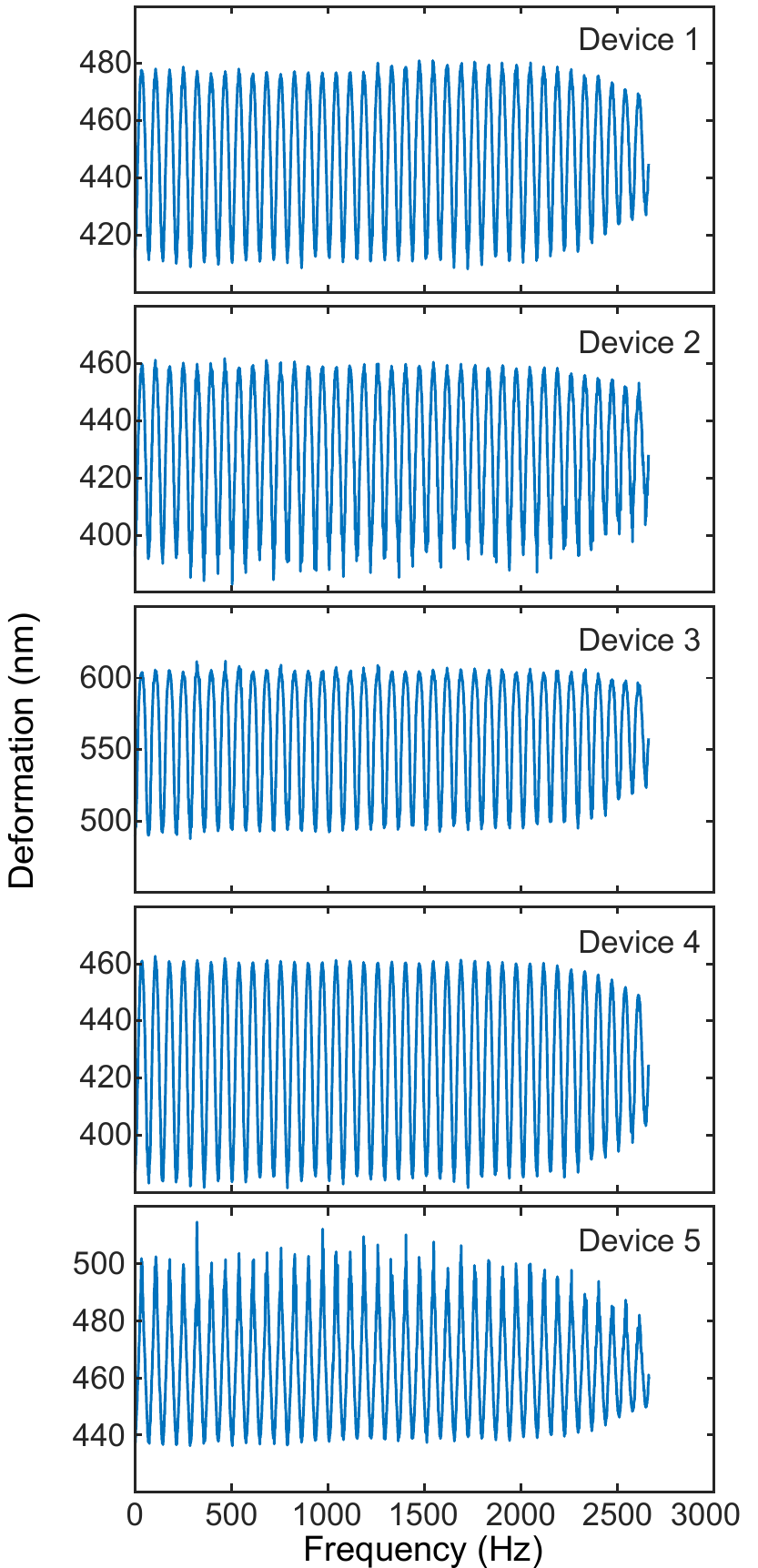}
	\caption{Displacement as extracted from stroboscopic measurement shown as a function of frequency for all devices. Note the larger response seen in Device~3.}
	\label{fgr:FigS6}
\end{figure}

\begin{figure}[H]
	\centering
	\graphicspath{{Figures/Supplementary_Info/}}
	\includegraphics[width=160mm]{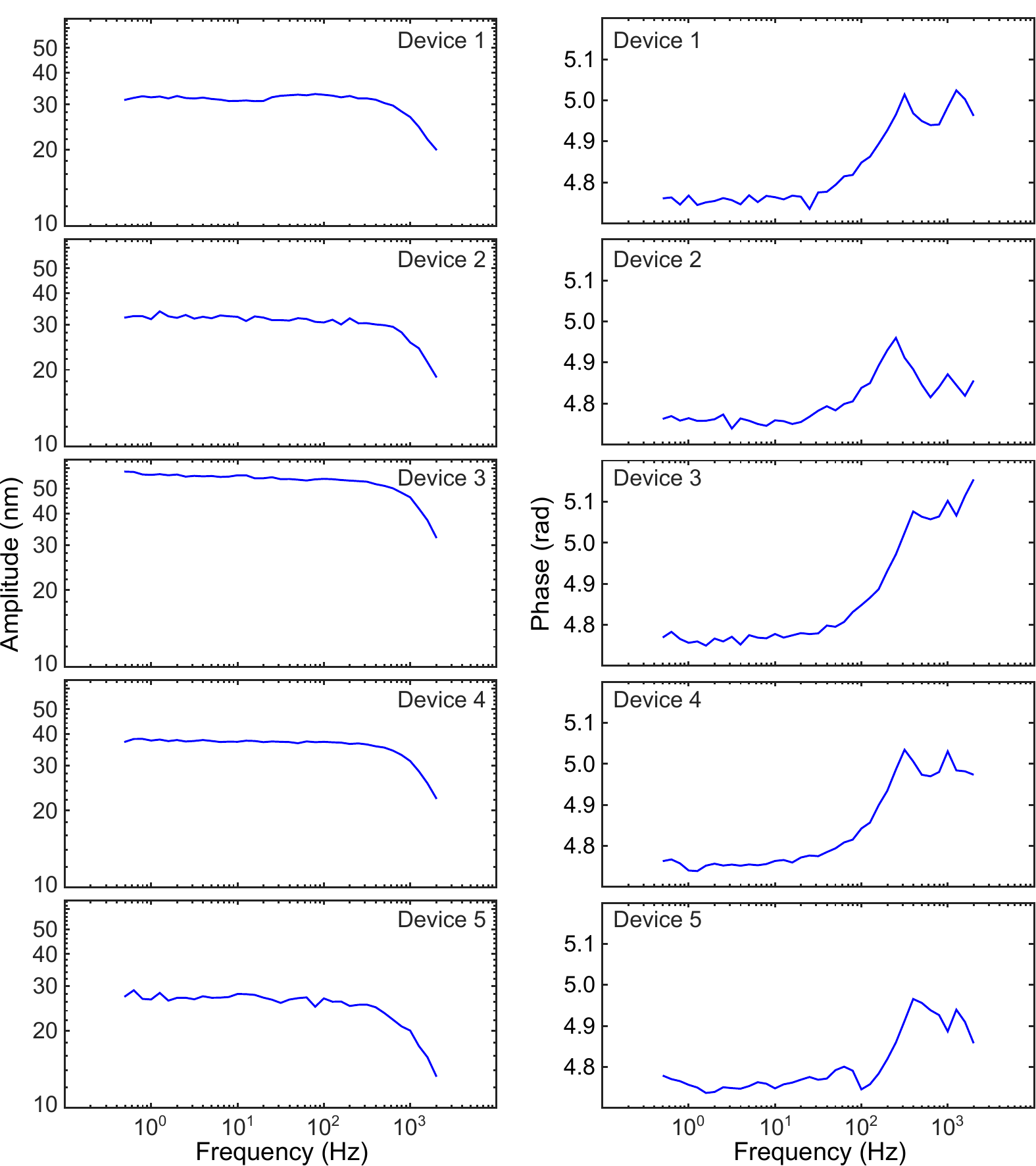}
	\caption{Phase and amplitude Bode plots for all strobscopically measured devices, showing that the response is flat up to 400~Hz (cutoff frequency of amplifier).}
	\label{fgr:FigS7}
\end{figure}

\nointerlineskip
Figures~\ref{fgr:FigS5},~\ref{fgr:FigS6} and \ref{fgr:FigS7} present the results of stroboscopic characterization for the all the measured devices. The device shown in the main text corresponds to Device 4 in the figures.

\newpage

\newpage
\section*{7. Simulating the RGB response of GIMOD pixels} 
\begin{figure}[H]
	\centering
	\graphicspath{{Figures/Supplementary_Info/}}
	\includegraphics[width=84.6mm]{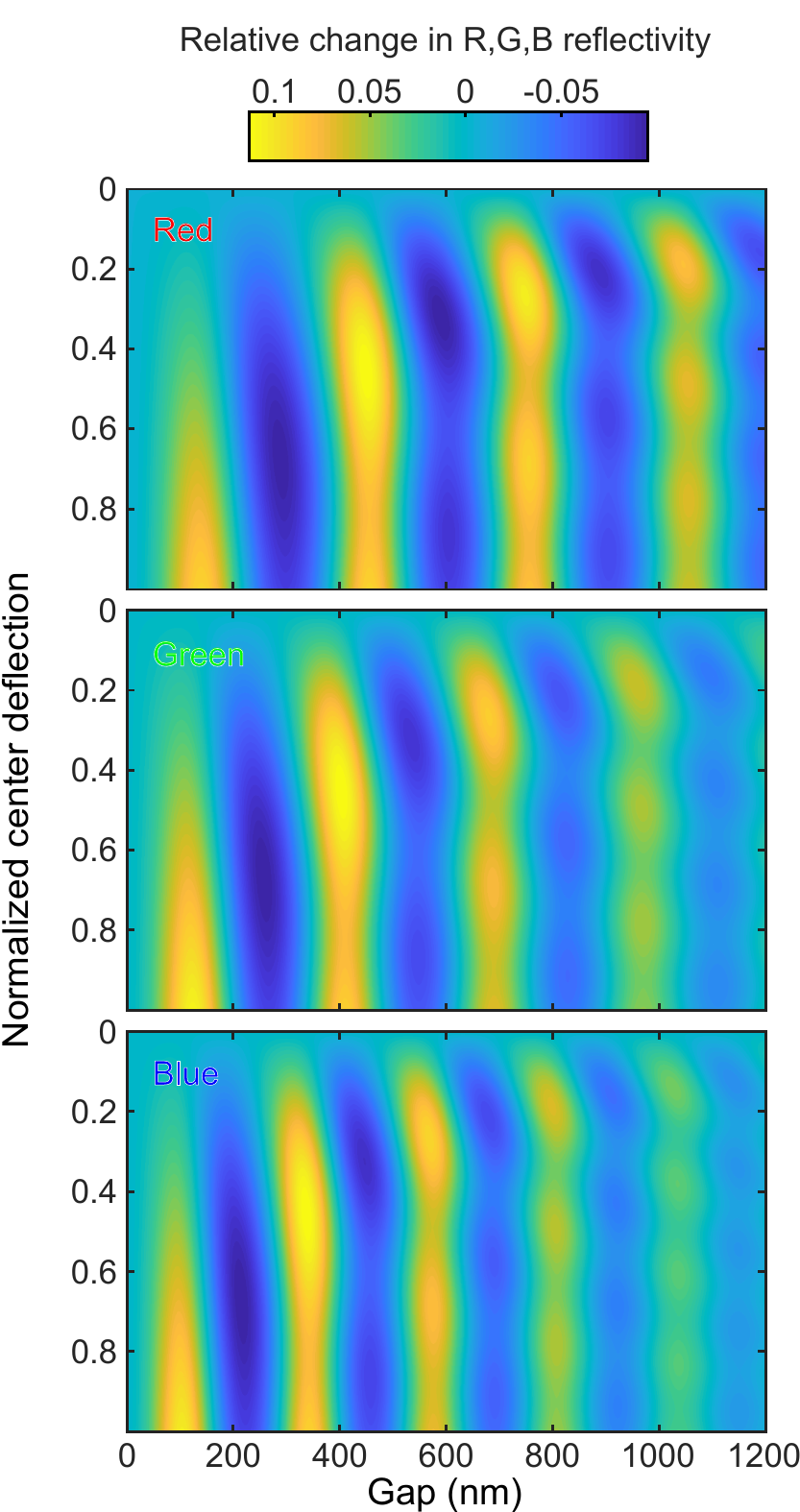}
	\caption{Simulated relative change in the R, G, B indices for a GIMOD pixel of 5~$\mu$m in diameter, under halogen lamp illumination.}
	\label{fgr:FigS8}
\end{figure}

\begin{figure}[H]
	\centering
	\graphicspath{{Figures/Supplementary_Info/}}
	\includegraphics[width=167.8mm]{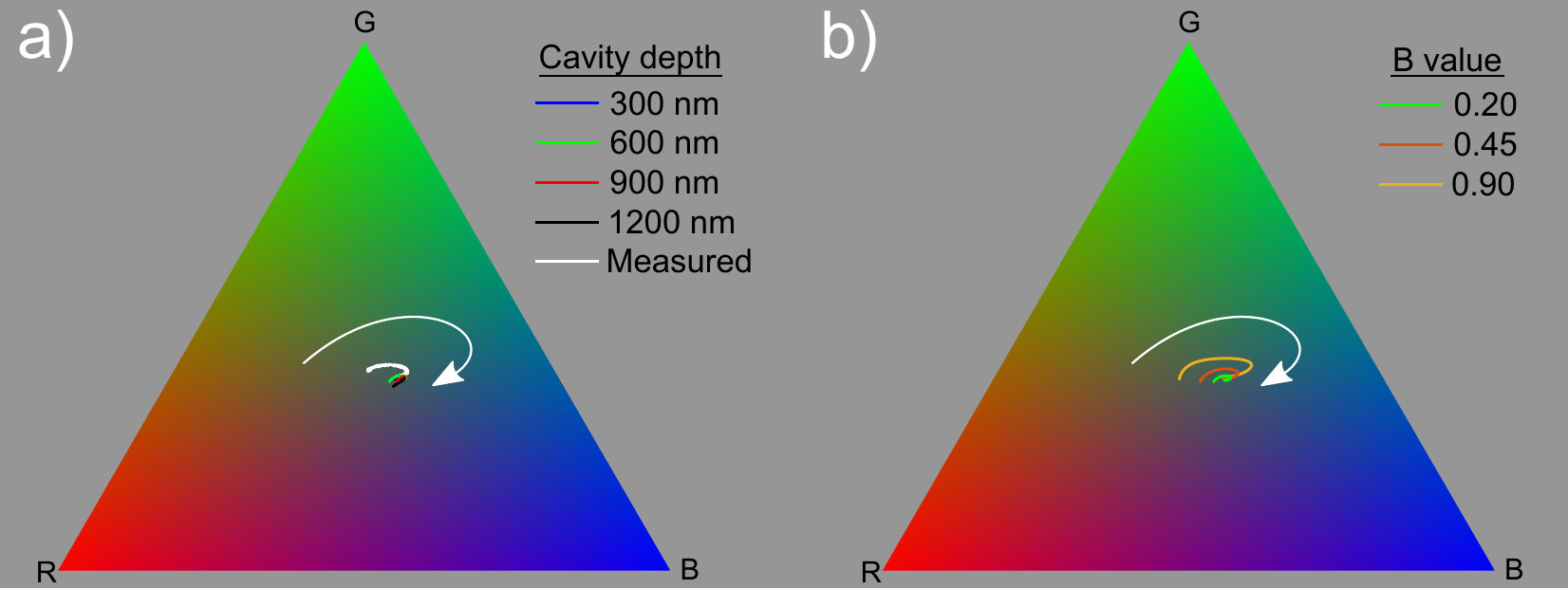}
	\caption{Simulated gamut pixel trajectory on a sRGB color triangle (CIE1931 colorimetry). a) Simulation of the average reflectance of a GIMOD pixel (5~$\mu$m in diameter) for different depths of the cavity (A=1, B=0.2). b) Simulation of the pixel trajectory for cavity depth of 600 nm, halogen illumination, and different values of B (A=1).}
	\label{fgr:FigS9}
\end{figure}

\nointerlineskip
The RGB response of a GIMOD pixel can be obtained using the following equation:
\begin{equation}
Red, Green, Blue =  \int_{0}^{\infty}{I(\lambda)C_{Red, Green, Blue}(\lambda)R_{avg}(\lambda)d\lambda},
\label{eqn:Eq11}
\end{equation}
\noindent
where $R_{avg}(\lambda)$ is the wavelength dependent drum average reflectivity obtained in equation~\ref{eqn:Eq10}, $I(\lambda)$ is the power spectrum of the illumination source, and $C_{Red, Green, Blue}$ is the CIE 1931 color matching functions for Red, Green, and Blue respectively [S-3]. The left hand side simply indicates the value for the Red, Green, or Blue.\\
The values obtained from equation~\ref{eqn:Eq11}, still need to undergo a gamma compression to account for the way human vision perceives colors. This is done by applying the following transformation:
\begin{equation}
RGB_{Corrected} =  RGB^{\gamma}\\
\label{eqn:Eq12}
\end{equation}
\noindent
In this work, the value for gamma in equation~\ref{eqn:Eq12} is taken to be $\gamma = \frac{1}{2.4}$, while the illumination profile $I(\lambda)$ used in these simulations is that of a halogen lamp [S-4]. The Red, Green, and Blue components for a GIMOD pixel are obtained numerically for various gaps and for A, B values fitted experimentally (A=0.32, B=0.063), the relative change in each color component as a function of center deflection and gap size is shown in Figure~\ref{fgr:FigS8}. \\
In order to plot these values as trajectories on an RGB color triangle, the following coordinate transformation is applied [S-3]:\\
\begin{equation}
\begin{split}
x = \frac{Blue}{Red + Green + Blue}\\
y = \frac{Green}{Red + Green + Blue}\\
\label{eqn:Eq13}
\end{split}
\end{equation}
Various trajectories of these GIMOD pixels on an RGB color triangle are equally plotted in Figure~\ref{fgr:FigS9}.

\newpage
\section*{References}

[S-1] Cartamil-Bueno, S.~J. {et~al.}. Colorimetry Technique for Scalable Characterization of Suspended Graphene. \textit{Nano Letters} 2016; 16:6792--6796.\\
\noindent
[S-2] Reserbat-Plantey, A. {et~al}. Electromechanical control of nitrogen-vacancy defect emission using graphene nems. \textit{Nature Communications} (2016).\\
\noindent
[S-3] Westland, S., Ripamonti, C., and Cheung, V. Computational colour science using MATLAB. \textit{John Wiley \& Sons} 2012.\\
\noindent
[S-4] \url{https://www.thorlabs.com/newgrouppage9.cfm?objectgroup_id=7541}

\clearpage